\documentclass[twocolumn,showpacs,preprintnumbers]{revtex4}
\usepackage{graphicx}
\usepackage{dcolumn}
\usepackage{bm}
\usepackage{subfigure}
\begin{document}
\newcommand{\ds}{\displaystyle}
\newcommand{\be}{\begin{equation}}
\newcommand{\en}{\end{equation}}
\newcommand{\bea}{\begin{eqnarray}}
\newcommand{\ena}{\end{eqnarray}}
\title{Morris-Thorne wormholes in static pseudo-spherically symmetric spacetimes}
\author{Mauricio Cataldo}
\altaffiliation{mcataldo@ubiobio.cl} \affiliation{Departamento de
F\'\i sica, Facultad de Ciencias, Universidad del B\'\i o-B\'\i o,
Avenida Collao 1202, Casilla 15-C, Concepci\'on, Chile.}
\author{Luis Liempi}
\altaffiliation{luiliempi@udec.cl} \affiliation{Departamento de
F\'\i sica, Universidad de Concepci\'on, Casilla 160-C,
Concepci\'on, Chile; and\\  Facultad de Ingenier\'\i a y
Tecnolog\'\i a, Universidad San Sebasti\'an, Campus Las Tres
Pascualas, Lientur 1457, Concepci\'on, Chile.}
\author{Pablo Rodr\'\i guez}
\altaffiliation{pablrodriguez@udec.cl} \affiliation{Departamento de
F\'\i sica, Universidad de Concepci\'on, Casilla 160-C,
Concepci\'on, Chile.\\}
\date{\today}
\begin{abstract}
In this paper we study classical general relativistic static
wormhole configurations with pseudo-spherical symmetry. We show that
in addition to the hyperbolic wormhole solutions discussed by Lobo
and Mimoso in the Ref. Phys.\ Rev.\ D {\bf 82}, 044034 (2010), there
exists another wormhole class, which is truly pseudo-spherical
counterpart of spherical Morris-Thorne wormhole (contrary to the
Lobo-Mimoso wormhole class), since all constraints originally
defined by Morris and Thorne for spherically symmetric wormholes are
satisfied. We show that, for both classes of hyperbolic wormholes
the energy density, at the throat, is always negative, while the
radial pressure is positive, contrary to the spherically symmetric
Morris-Thorne wormhole. Specific hyperbolic wormholes are
constructed and discussed by imposing different conditions for the
radial and lateral pressures, or by considering restricted choices
for the redshift and the shape functions. In particular, we show
that an hyperbolic wormhole can not be sustained at the throat by
phantom energy, and that there are pseudo-spherically symmetric
wormholes supported by matter with isotropic pressure and
characterized by space sections with an angle deficit (or excess).

\vspace{0.5cm} 
\end{abstract}
\smallskip
\maketitle 

\section{Introduction}
In classical general relativity the Einstein field equations admit a
simple and interesting class of static solutions describing tunnels
in spacetime, connecting either two remote regions of our Universe
or even different universes. Far from the tunnel, spacetime may
either be flat or curved geometry. These geometrical configurations
are called wormholes.

The spacetime wormhole ansatz of Morris and Thorne was formulated
originally for static spherically symmetric metrics in the
form~\cite{Morris:1988cz,Morris:1988tu}
\begin{eqnarray}\label{general BH metric}
ds^2=e^{2 \phi(r)} dt^2- \frac{dr^2}{1-\frac{b(r)}{r}}-r^2 \left(
d\theta^2+\sin^2\theta d\phi^2 \right),
\end{eqnarray}
where $e^{\phi(r)}$ and $b(r)$ are arbitrary functions of the radial
coordinate. In general, any static spherically symmetric spacetime
may be written in the form of Eq.~(\ref{general BH metric}). In the
case of wormhole configurations $e^{\phi(r)}$ and $b(r)$ are
referred to as redshift function and shape function respectively. In
order to have a wormhole these two functions must satisfy some
general constraints discussed by the authors in
Ref.~\cite{Morris:1988cz,Morris:1988tu}. These constraints provide a
minimum set of conditions which lead to a geometry featuring two
regions connected by a bridge, and are defined by:

Constraint 1: A no-horizon condition, i.e. $e^{\phi(r)}$ is finite
throughout the space–time in order to ensure the absence of horizons
and singularities.

Constraint 2: Minimum value of the $r$-coordinate, i.e. at the
throat of the wormhole $r = b(r) = r_0$, $r_0$ being the minimum
value of $r$.

Constraint 3: Finiteness of the proper radial distance, i.e.
\begin{eqnarray}\label{br1}
\frac{b(r)}{r} \leq 1,
\end{eqnarray}
(for $r \geq r_0$) throughout the space–time. The equality sign
holds only at the throat. Eq.~(\ref{br1}) is required in order to
ensure the finiteness of the proper radial distance $l(r)$ defined
by
\begin{eqnarray}
l(r) = \pm \int^r_{r_0} \frac{dr}{\sqrt{1-\frac{b(r)}{r}}},
\end{eqnarray}
where the $\pm$ signs refer to the two asymptotically flat regions
which are connected by the wormhole. Note that the
condition~(\ref{br1}) assures that the metric component $g_{rr}$ in
Eq.~(\ref{general BH metric}) does not change its sign for any $r
\geq r_0$.

Constraint 4: Asymptotic flatness condition, i.e. as $l \rightarrow
\pm \infty$ (or equivalently, $r \rightarrow \infty$) then
\begin{eqnarray}\label{MT C4}
\frac{b(r)}{r} \rightarrow 0.
\end{eqnarray}
Although asymptotically flat wormhole geometries are of particular
interest more general wormhole spacetimes also have been studied in
the literature~\cite{WH generales}.

Wormholes are sustained by exotic matter sources which would violate
all the energy conditions~\cite{Morris:1988cz,Morris:1988tu}. Static
wormholes supported by phantom energy also have been
constructed~\cite{Sushkov:2005kj,Lobo}. In this case the notion of
phantom energy is used in a more general sense than it has been used
in standard cosmology, where sources are characterized by an
isotropic pressure. For static spacetimes the phantom matter
threading the wormholes is essentially an inhomogeneous and
anisotropic fluid with radial and lateral pressures satisfying the
relation $p_r \neq p_l$. This inhomogeneous and anisotropic phantom
matter has a very strong negative radial pressure with the equation
of state $p_r/\rho <-1$ (note that this implies that the energy
density is positive).

It is remarkable that wormhole configurations may be constructed
without needing any form of exotic matter sources in the framework
of alternative theories of gravity, such as Einstein-Gauss-Bonnet
theory, Lovelock models, Brans-Dicke theory, among
others~\cite{Mehdizadeh}.

In this paper we examine the extension of spherically symmetric
Morris and Thorne wormholes to pseudo-spherically symmetric
Lorentzian spacetimes, i.e. to wormholes with negatively curved
spatial sections satisfying the four constraints enumerated above.

The first analysis of static pseudo-spherically symmetric wormholes
was made in Ref.~\cite{Lobo HWH}, where the authors explored
physical properties and characteristics of these hyperbolic
solutions, and analyzed some specific static wormhole solutions
supported by exotic matter. Lobo and Mimoso study a class of
hyperbolic wormholes which differ radically from the Morris-Thorne
wormholes. In particular, this type of wormholes does not satisfy
the condition~(\ref{br1}), on the other hand, for example, it was
shown that at the throat the energy density of the material
threading the hyperbolic spacetime tunnel is always negative, while
the radial pressure is positive, contrary to the spherically
symmetric Morris-Thorne counterpart.

In this work we further explore these pseudo-spherically symmetric
wormholes. In particular, we show that there exist another class of
hyperbolic wormholes that satisfy all constraints summarized and
enumerated above, which were originally formulated by Morris and
Thorne (in the following M-T).

In order to compare both classes of pseudo-spherical wormholes we
summarize the constraints of traversable hyperbolic tunnels
formulated by Lobo and Mimoso (in the following L-M). The spacetime
wormhole ansatz of Lobo and Mimoso is written in the form~\cite{Lobo
HWH}
\begin{eqnarray}\label{LM BH metric}
ds^2=e^{2 \phi(r)} dt^2- \frac{dr^2}{\frac{b(r)}{r}-1}-r^2 \left(
d\theta^2+\sinh^2\theta d\phi^2 \right),
\end{eqnarray}
where $e^{\phi(r)}$ and $b(r)$ are the redshift and the shape
functions respectively.

The Constraint 1, on the absence of horizons, and Constraint 2, on
the minimum value of the $r$-coordinate at the throat of the
wormhole, are the same, implying that we must require that
$e^{\phi(r)}$ is finite throughout the spacetime, and that
$b(r_0)=r_0$, respectively.

For the Constraint 3 the Eq.~(\ref{br1}) is not more fulfilled by
hyperbolic L-M wormholes. The metric~(\ref{LM BH metric}) imposes
the condition
\begin{eqnarray}\label{condition lm}
\frac{b(r)}{r}  \geq 1,
\end{eqnarray}
for having $g_{rr} >0$ for $r \geq r_0$.

The current formulation of the M-T Constraint 4, on asymptotic
flatness condition, clearly is not fulfilled by the metric~(\ref{LM
BH metric}), since Eq.~(\ref{condition lm}) implies that, as $r
\rightarrow \infty$, the relation~(\ref{MT C4}) can not be fulfilled
at all (note also that if $b(r)/r \rightarrow 0$, then the radial
metric component in Eq.~(\ref{LM BH metric}) becomes negative).

Notice that Lobo and Mimoso consider that the constructed static and
pseudo-spherically symmetric spacetime tunnels are made by adding
exotic matter to the vacuum solution $e^{\phi(r)}=2\mu/r-1$,
$b(r)=2\mu$, which is the pseudo-spherically counterpart of the
vacuum Schwarzschild solution. We shall address this statement in
the conclusion section.

The paper is organized as follows. In Sec. II we write the Einstein
equations for static pseudo-spherically symmetric spacetimes. In
Sec. III we analyze hyperbolic wormholes sustained by a matter
source with anisotropic pressure, characterized by a linear equation
of state for the radial pressure, and obtain hyperbolic counterparts
of spherical zero-tidal-force wormholes. In Sec. IV we analyze
pseudo-spherical wormhole models sustained by a fluid with isotropic
pressure. In Section V we discuss our results.

\section{Field equations for static pseudo-spherically symmetric spacetimes}
In order to describe hyperbolic M-T wormhole class we shall use the
wormhole ansatz in the form
\begin{eqnarray}\label{general pseudospherically BH metric}
ds^2=e^{2 \phi(r)} dt^2- \frac{dr^2}{1-\frac{b(r)}{r}}-r^2
d\Omega_{ps}^{2},
\end{eqnarray}
where $d\Omega_{ps}^{2}= d\theta^2+\sinh^2\theta d\phi^2$,
$e^{\phi(r)}$ and $b(r)$ are arbitrary functions of the radial
coordinate, and the usual two-dimensional spheres
$d\theta^2+\sin^2\theta d\phi^2$ of the metric~(\ref{general BH
metric}) are replaced by two-dimensional pseudo-spheres
$d\theta^2+\sinh^2\theta d\phi^2$. The coordinates $\theta$ and
$\phi$ in this case are defined in the ranges $-\infty < \theta <
\infty$ and $0\leq \phi \leq 2 \pi$ respectively. In general, any
static pseudo-spherically symmetric spacetime may be written in the
form~(\ref{LM BH metric}) or~(\ref{general pseudospherically BH
metric}). The essential characteristics of a wormhole geometry are
still encoded in the spacelike section, so as in spherically
symmetric wormholes, we shall call $e^{\phi(r)}$ and $b(r)$ the
redshift and shape functions respectively.

The L-M metric~(\ref{LM BH metric}) differs from Eq.~(\ref{general
pseudospherically BH metric}) only by the metric component $g_{rr}$.
It becomes clear that if we impose conditions on the redshift
function or on the energy density and pressures (by requiring, for
example, an equation of state) the spacetimes generated by the
metric ansatz~(\ref{LM BH metric}) and~(\ref{general
pseudospherically BH metric}), and satisfying the Einstein
equations, will be the same solutions. However, if we construct
wormhole solutions by imposing specific restricted choices for the
shape function, the constraints imposed by the metrics~(\ref{LM BH
metric}) and~(\ref{general pseudospherically BH metric}) on $b(r)$
may be quite different. Effectively, if we consider for example the
shape function in the form $b(r)=r_0(r/r_0)^\alpha$,
the ansatz metric~(\ref{LM BH metric}), used in Ref.~\cite{Lobo
HWH}, implies that the radial metric component is given by
$g_{rr}^{-1}=\left( \frac{r}{r_0} \right)^{\alpha-1}-1$. Then, in
order to have $g_{rr}
>0$, for $r\geq r_0$, we must require that $\alpha>1$. On the other
hand, for the same form of the shape function we have that the
metric~(\ref{general pseudospherically BH metric}) implies that the
radial metric component is given by
$g_{rr}^{-1}=1-\left(\frac{r}{r_0} \right)^{\alpha-1}$. Then, in
order to have $g_{rr}
>0$, for $r\geq r_0$, we must require that $\alpha<1$. This statement
will be addressed below by considering this specific form of the
shape function $b(r)$.

By assuming that the matter content is described by a single
imperfect fluid, from the Einstein field equations we obtain
\begin{eqnarray}
\kappa \rho(r)=\frac{b^{\prime}-2}{r^2}, \label{p rho} \\
\kappa p_r(r)=2\left(1-\frac{b}{r}  \right)
\frac{\phi^{\prime}}{r}-\frac{b}{r^3}+\frac{2}{r^2}, \label{p pr} \\
\kappa p_l(r)=\left( 1-\frac{b}{r} \right) \times \nonumber
\\ \left[\phi^{\prime \prime}+\phi^{\prime \, 2} - \frac{b^\prime
r+b-2r}{2r(r-b)} \, \phi^\prime - \frac{b^\prime
r-b}{2r^2(r-b)}\right], \label{p pl}
\end{eqnarray}
where $\kappa=8 \pi G$, prime denotes $\partial/\partial r$, $\rho$
is the energy density, and $p_r$ and $p_l$ are the radial and
lateral pressures respectively.

From Eqs.~(\ref{p rho})-(\ref{p pl}) we obtain the hydrostatic
equation for equilibrium of the matter content sustaining the
hyperbolic wormhole geometry, which is given by
\begin{eqnarray}\label{conservation equation sinh}
p_r^\prime=\frac{2(p_l-p_r)}{r}- (\rho+p_r) \phi^\prime.
\end{eqnarray}
Summarizing, we  have a system of three equations~(\ref{p
rho})-(\ref{p pl}), with five unknown functions: $e^{\phi(r)}$,
$b(r)$, $\rho(r)$, $p_r(r)$, and $p_l(r)$. To construct hyperbolic
wormholes we must constraint two of these five functions. In order
to make this we will consider specific equations of state for $p_r$
or $p_l$, impose the condition $p_r=p_l$, for having isotropic
pressure, or consider restricted choices for the redshift and shape
functions.

\section{Hyperbolic wormholes with $p_r=\omega \rho$}
Let us begin our study by considering pseudo-spherical wormhole
solutions sustained by a matter with anisotropic pressure,
characterized by the linear equation of state
\begin{eqnarray}\label{promegarho}
p_r=\omega \rho,
\end{eqnarray}
for the radial pressure. By using Eqs.~(\ref{p rho}), (\ref{p pr})
and~(\ref{promegarho}) we obtain the following equation
\begin{eqnarray}\label{promegarho final}
\omega r b^\prime+2 r (b-r) \phi^\prime +b=2(1+\omega)r.
\end{eqnarray}
For zero-tidal-force wormholes, i.e. $\phi(r)=0$, we obtain
\begin{eqnarray}
b(r)=2r+Cr^{-1/\omega},
\end{eqnarray}
where $C$ is an integration constant, and then we may write the
solution in the form
\begin{eqnarray}\label{metricalobo}
ds^2&=&  dt^2- \frac{dr^2}{\left(\frac{r}{r_0}
\right)^{-\frac{1+\omega}{\omega }}-1}-
 r^2 d\Omega_{ps}^{2}, \nonumber \\ \label{densidadlobo} \\
\kappa \rho&=&\frac{\left(\frac{r}{r_0} \right)^{-(3 \omega+1)/\omega}}{\omega r_0^2}, \\
\kappa p_l&=&-\frac{(1+\omega)\left(\frac{r}{r_0} \right)^{-(3
\omega+1)/\omega}}{2 \omega r_0^2}.\label{presionlobo}
\end{eqnarray}
In this case we have that for $r\geq r_0$ the metric component
$g^{-1}_{rr} \geq 0$ if $-1 \leq \omega<0$. This implies that the
energy density is negative, and the pressures $p_r$ and $p_l$ are
positive. This solution was previously obtained by the authors of
Ref.~\cite{Lobo HWH}. If we instead of the equation of
state~(\ref{promegarho}) require such a linear equation of state for
the lateral pressure, i.e. $p_l=\tilde{\omega} \rho$, then the same
solution is obtained. Both of them are connected by the
transformation $\tilde{\omega}=-(1+\omega)/2$.

Now we will consider the wormhole solution generated by imposing on
the metric~(\ref{general pseudospherically BH metric}) the shape
function in the form $b(r)=r_0(r/r_0)^\alpha$. From
Eq.~(\ref{promegarho final}) we obtain that
\begin{eqnarray}\label{ass15}
e^{\phi(r)}= \frac{A \left(r_0 \left(\frac{r}{r_0} \right)^\alpha
-r\right)^{\frac{1+2\omega-\alpha\omega}{2(\alpha-1)}}}{\left(
\frac{r}{r_0}\right)^{\frac{-1+\alpha(\omega+2)}{2(\alpha-1)}}},
\end{eqnarray}
where $A$ is an integration constant. Without any loss of generality
we can put $A=1$. Now, since Eq.~(\ref{ass15}) implies that
$e^{\phi(r_0)}=0$, an event horizon is located at $r_0$, implying
that in this case we can have only non-traversable hyperbolic
wormholes (including the case $\alpha=0$).

For avoiding this event horizon at $r_0$, we need to impose that
$\alpha=(1+2 \omega)/\omega$, then
$e^{\phi(r)}=\left(\frac{r}{r_0}\right)^{-\omega-1}$. Hence the
solution takes the form
\begin{eqnarray}\label{metrica15}
ds^2&=&\left(\frac{r}{r_0}\right)^{-2(\omega+1)} dt^2-
\frac{dr^2}{1-\left(\frac{r}{r_0}\right)^{\frac{\omega+1}{\omega}}}-r^2 d\Omega_{ps}^{2}, \nonumber \\ \\
\label{densidad15} \kappa
\rho&=&\frac{(2\omega+1)\left(\frac{r}{r_0}\right)^{\frac{1+\omega}{\omega}}-2
\omega}{\omega r^2},\\
\kappa p_l&=&-\frac{(1+\omega)\left(\left( \frac{r}{r_0}
\right)^{\frac{1+\omega}{\omega}}
(1+2\omega)-2(1+\omega)\right)}{2r^2}, \nonumber \\
\label{presion15}
\end{eqnarray}
where $\omega \neq -1$. In order to have a wormhole we must require
$-1< \omega< 0$. Note that at the throat the energy
density~(\ref{densidad15}) takes the value $\rho(r_0)=\frac{1}{r_0^2
\omega}$, which is negative for $\omega<0$. This matter
configuration violates the strong energy condition since
$\rho+p_r+2p_l<0$ at the throat. It is easy to see that for $-1<
\omega< 0$ and $r \geq r_0$ the energy density is always negative,
and $p_r>0$.

It is clear that the
wormhole~(\ref{metricalobo})-(\ref{presionlobo}) is quite different
from the wormhole~(\ref{metrica15})-(\ref{presion15}). For the first
one, we have that if $r \rightarrow \infty$ then $b(r)/r \rightarrow
-\infty$, while for the second wormhole $b(r)/r \rightarrow 0$. The
latter behavior is the same as of spherically symmetric M-T
wormholes, since the Eq.~(\ref{br1}) is satisfied. So we can say
that the wormhole~(\ref{metrica15})-(\ref{presion15}) is the
hyperbolic version of the respective spherically symmetric wormhole
\begin{eqnarray}\label{a15}
ds^2=\left(\frac{r}{r_0}\right)^{-2(\omega+1)} dt^2-
\frac{dr^2}{1-\left(\frac{r}{r_0}\right)^{\frac{\omega+1}{\omega}}}+
\nonumber \\ r^2 \left( d\theta^2+\sin^2\theta d\phi^2 \right),
\end{eqnarray}
which may be obtained by the replacement of $\sinh \theta$ by $\sin
\theta$. However, note that the spherical wormhole~(\ref{a15}) does
not fulfill the equation of state~(\ref{promegarho}).

Lastly, in this section, we will construct the Lorentzian hyperbolic
M-T wormhole by imposing to the metric~(\ref{general
pseudospherically BH metric}) the restricted choices $b(r)=r_0
(r/r_0)^\alpha$, for the shape function, and $\phi=0$, for the
redshift function. This wormhole is the counterpart of the
zero-tidal-force spherically symmetric Morris and Thorne wormhole
discussed in Ref.~\cite{Morris:1988cz}, which is given by
\begin{eqnarray}\label{ztf wh}
ds^2=dt^2- \frac{dr^2}{1-\left(\frac{r_0}{r} \right)^{1-\alpha}}-r^2
\left( d\theta^2+\sin^2\theta d\phi^2 \right),
\end{eqnarray}
where $\alpha<1$. In this case the energy density and pressures are
given by
\begin{eqnarray}
\kappa \rho= \frac{ \alpha  \left( \frac{r_0}{r}
\right)^{3-\alpha}}{r_0^2}, \\
\kappa p=- \frac{ \left( \frac{r_0}{r} \right)^{3-\alpha}}{r_0^2},
\\
\kappa p_l= \frac{ (1-\alpha)  \left( \frac{r_0}{r}
\right)^{3-\alpha}}{2r_0^2},
\end{eqnarray}
Note that this spherical spacetime is asymptotically flat and the
pressures satisfy linear equations of state. For $0<\alpha<1$ the
energy density is positive, while for $\alpha<0$ it becomes
negative.

Thus by putting $\phi=0$ and $b(r)=r_0 (r/r_0)^\alpha$ into the
field equations~(\ref{p rho})-(\ref{p pl}) we obtain that the
zero-tidal-force hyperbolic counterpart is given by
\begin{eqnarray}\label{ztf hwh}
ds^2=dt^2- \frac{dr^2}{1-\left(\frac{r_0}{r} \right)^{1-\alpha}}-r^2
d\Omega_{ps}^{2}, \\ \label{ed hmt wh}\kappa \rho= \frac{ \alpha
\left( \frac{r_0}{r}
\right)^{3-\alpha}}{r_0^2}-\frac{2}{r_0^2}\left(\frac{r_0}{r}\right)^2, \\
\kappa p_r=- \frac{ \left( \frac{r_0}{r}
\right)^{3-\alpha}}{r_0^2}+\frac{2}{r_0^2}\left(\frac{r_0}{r}\right)^2,
\\
\kappa p_l= \frac{ (1-\alpha)  \left( \frac{r_0}{r}
\right)^{3-\alpha}}{2r_0^2}, \label{plhwh}
\end{eqnarray}
where $\alpha<1$. It becomes clear that this zero-tidal-force
solution belongs to the hyperbolic M-T wormhole class, while the
zero-tidal-force wormhole~(\ref{metricalobo}) belongs to the L-M
wormhole class.

It can be shown that the energy density~(\ref{ed hmt wh}) is always
negative for $r \geq r_0$. The condition $\alpha<1$ implies that for
$r\approx 0$ the first term of Eq.~(\ref{ed hmt wh}) dominates over
the second one. Then, $\rho \geq 0$ for $0<r \leq r_0
(\frac{2}{\alpha})^{1/(\alpha-1)}$, and $\rho < 0$ for $r> r_0
(\frac{2}{\alpha})^{1/(\alpha-1)}$. However, for $\alpha<1$ always
$r_0 (\frac{2}{\alpha})^{1/(\alpha-1)} <r_0$, implying that the
energy density is always negative for $r\geq r_0$.

It is interesting to note that the wormholes~(\ref{ztf wh})
and~(\ref{ztf hwh}) satisfy the energy condition $\rho+p_r+2 p_l
\geq 0$, since for them we have that $\rho+p_r+2 p_l= 0$.

Both wormholes have the same embedding diagrams for slices
$t=const$, $\theta=\pi/2$ in the spherical metric~(\ref{ztf wh}),
and $t=const$, $\theta=\ln(1+\sqrt{2})$ in the pseudo-spherical
metric~(\ref{ztf hwh}).

Note that the background on which is constructed the spherical
wormhole~(\ref{ztf wh}) is the Minkowski spacetime, since for
$r_0=0$ we have $\rho=p_r=p_l=0$. On the other hand, the asymptotic
spacetime of the metric~(\ref{ztf wh}) is also the Minkowski space.

The background on which we construct the hyperbolic
wormhole~(\ref{ztf hwh}) is not an empty spacetime since for the
metric
\begin{eqnarray}\label{hip}
ds^2=dt^2- dr^2 -r^2 \left( d\theta^2+\sinh^2\theta d\phi^2 \right),
\end{eqnarray}
we have that
\begin{eqnarray}
\kappa \rho&=&-\kappa p_r =-\frac{2}{r^2},
\\
\kappa p_l &=& 0.
\end{eqnarray}
This spacetime is not flat, but it is asymptotically locally flat,
since for $r \rightarrow \infty$ we obtain that $\rho \rightarrow
0$, $p_r \rightarrow 0$, and for observers, who always remain at
rest at constant $r$, $\theta$ and $\phi$, we obtain that, at the
proper orthonormal basis, the curvature $R_{(\alpha) (\beta)
(\gamma) (\delta)} \rightarrow 0$ if $r \rightarrow \infty$.

Thus the interpretation is direct: we have constructed a static
wormhole on the non-empty and asymptotically locally flat
background~(\ref{hip}) by adding the exotic matter source given
by~(\ref{ed hmt wh})-(\ref{plhwh}) to it. Notice that this is not
true for all hyperbolic wormhole spacetimes. See, for example, the
wormhole metrics~(\ref{metricalobo}) or~(\ref{WH prho3}) below: the
backgrounds on which these wormholes are constructed and their
asymptotic spacetimes do not have the form of the
metric~(\ref{hip}).

\section{Pseudo-spherical wormholes with isotropic pressure}
Now, in this section, we shall construct new pseudo-spherical
wormholes sustained by a matter content with isotropic pressure. The
main condition for having an isotropic pressure is to require for
the radial and lateral pressures the constraint $p_r=p_l$. Thus, for
hyperbolic spacetimes, from Eqs.~(\ref{p pr}) and~(\ref{p pl}), we
have that equation
\begin{eqnarray}\label{p isotropic condition}
\phi^{\prime \prime}+\phi^{\prime \, 2} - \frac{b^\prime
r-3b+2r}{2r(r-b)} \, \phi^\prime = \frac{b^\prime
r-3b+4r}{2r^2(r-b)}
\end{eqnarray}
must be fulfilled by the metric functions $\phi(r)$ and $b(r)$. From
this equation we get for the shape function
\begin{widetext}
\begin{eqnarray}\label{isotropic integral 15}
b(r)=\left ( \int\frac{2 \left( r^2 \phi^{\prime \prime}+r^2
\phi^{\prime 2}-r \phi^\prime -2 \right) e^{\int
\frac{2r^2\phi^{\prime \prime}+2r^2\phi^{\prime
2}-3r\phi^\prime-3}{r(1+r \phi^\prime)}dr}}{1+r \phi^\prime}\, dr
+C\right) \, e^{-\int \frac{2r^2\phi^{\prime
\prime}+2r^2\phi^{\prime 2}-3r\phi^\prime-3}{r(1+r \phi^\prime)}
dr},
\end{eqnarray}
\end{widetext}
where $C$ is an integration constant. Equations~(\ref{p isotropic
condition}) and~(\ref{isotropic integral 15}) have a general
character since they do not involve an equation of state for the
energy density $\rho$ and the isotropic pressure $p$. Now we have
three differential equations for four unknown functions, namely
$\phi(r)$, $b(r)$, $\rho(r)$ and $p(r)$. Thus, to study solutions to
these field equations, we shall consider restricted choices for one
of the unknown functions.

For zero-tidal-force wormhole configurations the restricted form
$\phi=\phi_0=const$ must be required. By putting $\phi(r)=const$
into Eq.~(\ref{isotropic integral 15}) we obtain a spacetime of
constant curvature defined by the shape function $b(r)=2 r+Cr^3$. In
this case the metric, the pressure and the energy density may be
written in the form
\begin{eqnarray}\label{WH prho3}
ds^2= dt^2- \frac{dr^2}{\left(\frac{r}{r_0}\right)^2-1 }-r^2 \left(
d\theta^2+\sinh^2\theta d\phi^2 \right),
\end{eqnarray}
$p=-\rho/3=\frac{1}{\kappa r_0^2}$, respectively. Note that the
pressure satisfies a barotropic equation of state with a state
parameter $\omega=-1/3$, describing a gas of strings. In a
Friedmann-Robertson-Walker universe this gas of strings can change
the observable topology of the space~\cite{Kamenshchik:2011jq},
while in static spherically symmetric spacetimes it presence
produces rather bizarre geometries and it may influence on the
rotation curves, mimicking the dark matter
effects~\cite{Kamenshchik:2011jq}.

The metric~(\ref{WH prho3}) describes a wormhole, which belongs to
the family of L-M solutions~(\ref{metricalobo}). The throat is
located at $r=r_0$, the energy density is negative everywhere, and
the pressure is always positive. This hyperbolic wormhole is the
static version of the evolving wormhole discussed in Ref.~\cite{Li},
which represents a spacetime of two open Friedmann-Robertson-Walker
universes connected by a Lorentzian wormhole. The metric~(\ref{WH
prho3}) is obtained from the evolving wormhole studied in
Ref.~\cite{Li} by making $a(t)=const$.

Now, we shall construct a master equation for the shape function
$b(r)$, by using Eqs.~(\ref{p rho}),~(\ref{conservation equation
sinh}) and ~(\ref{p isotropic condition}) and assuming the linear
equation of state $p_r=p_l=\omega \rho$. The result of these
manipulations provides the following equation:
\begin{widetext}
\begin{eqnarray}\label{master eq}
\omega\left( 1+\omega\right) b'''-\frac{2\left(
\frac{1}{2}+\omega\right)\omega b''^{2}
}{b'-2}-\frac{\omega\left(2r-6\omega r +
\left(5\omega-3 \right)b+r\left(1+\omega \right)b' \right)b''  }{2r(r-b)}+ \nonumber \\
\frac{3\left(\frac{4}{3}r
(\omega^{2}+1+4\omega)-(1+\frac{16}{3}\omega+\frac{5}{3}\omega^{2})b
+r(\frac{1}{3}+\omega)(1+\omega)b'\right)\left(b'-2 \right)
}{2r^{2}(r-b)}=0.
\end{eqnarray}
\end{widetext}

It can be shown that the previous discussed solution~(\ref{WH
prho3}) satisfies the master equation~(\ref{master eq}) identically
as well as $b(r)=2 r-2\mu$, where $\mu$ is a constant. The latter
implies that
\begin{eqnarray}\label{VSS}
ds^2=\left( \frac{2 \mu}{r}-1  \right) dt^2- \frac{dr^2}{\frac{2
\mu}{r}-1}-r^2 d\Omega_{ps}^{2},
\end{eqnarray}
with $\rho=p_r=p_l=0$, and represents the hyperbolic counterpart of
the spherically symmetric vacuum Schwarzschild solution. It must be
noticed that the metric~(\ref{VSS}) is static only in the region $0
< r <2\mu$, therefore this hyperbolic solution does not represent
neither a black hole nor a non-traversable wormhole (for a
discussion about the spacetime~(\ref{VSS}) see~\cite{Lobo HWH} and
the references there in).

Another interesting solution to this master equation is given by
$b(r)=2r-A-\frac{1}{3} B r^3$, and the metric takes the form
\begin{eqnarray}\label{hyperbolic Kottler}
ds^2= \left(\frac{A}{r}+\frac{1}{3} B r^2-1 \right) dt^2-
\frac{dr^2}{\frac{A}{r}+\frac{1}{3} B r^2-1 }-r^2 d\Omega_{ps}^{2},
\nonumber \\
\end{eqnarray}
where $p=-\rho=B$. This metric is the hyperbolic version of the
spherically symmetric Kottler solution~\cite{Kottler}. It is
interesting to note that this solution may represents a black hole
spacetime for Einstein equations with a cosmological
constant~\cite{Cai Wang15}, and with an appropriate choice of
parameter values the spacetime may have a single event horizon.

For $A=0$ the spacetime~(\ref{hyperbolic Kottler}) becomes a
non-traversable hyperbolic wormhole.

\subsection{Hyperbolic wormhole solutions with isotropic
pressure for $e^{\phi(r)}=\left(\frac{r}{r_0}  \right)^\beta$}

Now we shall impose a restricted form for the redshift function. It
is of interest to study hyperbolic wormhole solutions generated by
the redshift function
\begin{eqnarray}\label{ephi}
e^{\phi(r)}=\left(\frac{r}{r_0} \right)^\beta,
\end{eqnarray}
with $\beta$ a constant. In the case of spherically symmetric
wormholes, such a redshift function generates wormhole spacetimes
which are not asymptotically flat, being that in the
metric~(\ref{general pseudospherically BH metric}) $g_{tt}
\rightarrow \infty$ for $r \rightarrow \infty$ and $\beta
>0$, and $g_{tt} \rightarrow 0$ for $r \rightarrow \infty$ and
$\beta <0$.

By putting Eq.~(\ref{ephi}) into Eq.~(\ref{p isotropic condition})
we obtain
\begin{eqnarray}
b(r)= \frac{\beta^2-2\beta-2}{\beta^2-2\beta-1} \, r+C
r^{-\frac{(2\beta+1)(\beta-3)}{1+\beta}},
\end{eqnarray}
where $C$ is an integration constant. Then the metric is given by
\begin{eqnarray}\label{2 metric 1AA}
ds^2=\left(\frac{r}{r_0} \right)^{2 \beta} dt^2- \nonumber \\
\frac{dr^2}{1-{\frac {\beta^2- 2\beta-2 }{\beta^2-2\,\beta-1}}-C \,
{\left(\frac{r}{r_0} \right)}^{-{\frac {2(\beta^2-2
\beta-1)}{1+\beta}}}}-  r^2 d \Omega_{ps}^2.
\end{eqnarray}
This metric  in the asymptotic limit $r \rightarrow \infty$
describes an hyperbolic spacetime carrying a topological defect for
${\frac {(\beta^2-2 \beta-1)}{1+\beta}}>0$ and $C>0$, implying that
the parameter $\beta$ varies in the ranges $-1<\beta< 1-\sqrt{2}$ or
$1+\sqrt{2}<\beta$. Thus, if $r \rightarrow \infty$ then $b(r)/r
\rightarrow \frac{\beta^2-2\beta-2}{\beta^2-2\beta-1}$, and the
asymptotic spacetime is given by
\begin{eqnarray*}\label{metric 15 DA}
ds^2=\left(\frac{r}{r_0} \right)^{2 \beta} dt^2-
({\beta}^{2}-2\,\beta-1) dr^2- r^2 d\Omega_{ps}^2,
\end{eqnarray*}
which, by making $(\beta^{2}-2\,\beta-1) \, r^2=  \varrho^2$, may be
rewritten as
\begin{eqnarray}\label{metric 15 DA 15}
ds^2=\left(\frac{\varrho}{\varrho_0} \right)^{2 \beta}
dt^2-d\varrho^2-({\beta}^{2}-2\,\beta-1 )^{-1} \varrho^2
d\Omega_{ps}^2.
\end{eqnarray}
This metric describes a space with an angle deficit (or excess) and
it may be interpreted as an hyperbolic spacetime carrying a
topological defect. Such an asymptotic spacetime with a topological
defect does not exist for the metric~(\ref{2 metric 1AA}) fulfilling
the constraints ${\frac {(\beta^2-2 \beta-1)}{1+\beta}}<0$ and
$C<0$. In this case we obtain that $g_{rr}^{-1} \rightarrow |C| \,
(r/r_0)^{\left|2{\frac {(\beta^2-2 \beta-1)}{1+\beta}}\right|}$ for
$r \rightarrow \infty$.

It must be noticed that, we use the term ``angle deficit (or
excess)", and not the term ``solid angle deficit (or excess)", as it
is used in the case of spherically symmetric spacetimes, since only
the coordinate $\phi$ describes an angle in the range $[0,2\pi]$.
From the metric~(\ref{metric 15 DA 15}) we see that for the
two-dimensional pseudo-spheres we have that $\alpha^2 \varrho^2
\left( d\theta^2+\sinh^2\theta d\phi^2 \right)$, where
$\alpha^2=({\beta}^{2}-2\,\beta-1 )^{-1}>0$. Thus, we see that the
two-dimensional pseudo-spherical part may be written as $\varrho^2
\left( d\tilde{\theta}^2+\sinh^2 (\tilde{\theta}/\alpha)
d\tilde{\phi}^2 \right)$, where $\tilde{\theta}=\alpha \theta$,
$\tilde{\phi}=\alpha \phi$ and $\alpha>0$, implying that $-\infty
<\tilde{\theta} < \infty$ and $0 \leq \tilde{\phi} \leq 2 \pi
\alpha$. It becomes clear that the new angle $\tilde{\phi}$ has an
angle deficit for $0<\alpha<1$, and an angle excess for $\alpha>1$.

Now, we shall show that the solution~(\ref{2 metric 1AA}) includes
hyperbolic L-M wormholes, as well as hyperbolic M-T ones.

\begin{widetext}

\begin{center}
\begingroup
\squeezetable
\begin{table}[h!]

\begin{tabular}{|c|c|c|c|c|c|} \hline
Metric & Constraints & Class & $\rho \geq 0$ & $\rho + p_{i} \geq 0$
& $\rho + p_{r} + 2p_{l} \geq 0$ \\ \hline
$ds^{2} = dt^{2} - \frac{dr^{2}}{\left( \frac{r}{r_{0}}\right)^{-\frac{1+\omega}{\omega}} -1} - r^{2}d\Omega_{ps}^{2}$, & $p_{r}=\omega \rho$, $\Phi(r)=0$,  & L-M & No & No & No \\
 & $-1<\omega <0$ & & & & \\ \hline
$ds^{2} = \left( \frac{r}{r_{0}} \right)^{-2\left( \omega +1 \right)} dt^{2} - \frac{dr^{2}}{1- \left(\frac{r}{r_{0}}\right)^{\frac{1+\omega}{\omega}}} - r^{2}d\Omega_{ps}^{2}$, & $p_{r}=\omega \rho$, & M-T & No & No & No \\
 & $-1<\omega <0$ & & & & \\ \hline
$ds^{2} = dt^{2} - \frac{dr^{2}}{1- \left( \frac{r_{0}}{r}\right)^{1-\alpha}} - r^{2}d\Omega_{ps}^{2}$, &   $\Phi(r)=0$, & M-T & No & No & Yes\\
 & $\alpha < 1$ & & & & \\ \hline
%
%
$ds^{2} = dt^{2} - \frac{dr^{2}}{\left(
\frac{r}{r_{0}}\right)^{2}-1} - r^{2}d\Omega_{ps}^{2}$ &
$p_{r}=p_{l}$, $\Phi(r)=0$ & L-M & No & No & No \\ \hline
%
%
%
%
$ds^{2} = \left( \frac{r}{r_{0}} \right)^{2\left( 1+ \sqrt{3}
\right)}dt^{2} - \frac{dr^{2}}{1- \left( \frac{r_{0}}{r}
\right)^{2/\left( 2 + \sqrt{3} \right)}} - r^{2}d\Omega_{ps}^{2}$ &
$p_{r}=p_{l}$, $e^{\Phi(r)}=\left(\frac{r}{r_{0}} \right)^{1 +
\sqrt{3}}$ & M-T & No & No & Yes \\ \hline
$ds^{2} = \left( \frac{r}{r_{0}} \right)^{2\left( 1- \sqrt{3}
\right)}dt^{2} - \frac{dr^{2}}{1- \left( \frac{r_{0}}{r}
\right)^{2/\left( 2 - \sqrt{3} \right)}} - r^{2}d\Omega_{ps}^{2}$ &
$p_{r}=p_{l}$, $e^{\Phi(r)}=\left(\frac{r}{r_{0}} \right)^{1 -
\sqrt{3}}$ & M-T & No & No & No \\ \hline
$ds^{2} = \left( \frac{r}{r_{0}} \right)^{4}dt^{2} -
\frac{dr^{2}}{\left( \frac{r}{r_{0}} \right)^{2/3} -1} -
r^{2}d\Omega_{ps}^{2}$ & $p_{r}=p_{l}$,
$e^{\Phi(r)}=\left(\frac{r}{r_{0}} \right)^{2}$ & L-M & No & No &
Yes \\ \hline
$ds^{2}=\left( \frac{\varrho}{\varrho_{0}} \right)^{2\beta}dt^{2} - \frac{d\varrho^{2}}{\left( \frac{\varrho}{\varrho_{0}} \right)^{-\frac{2\left( \beta^{2}-2\beta -1 \right)}{1+\beta}}  -1} -\left( \frac{\beta^{2}-2\beta -2}{\beta^{2}-2\beta -1} -1 \right)\varrho^{2}d\Omega_{ps}^{2}$, & $p_{r}=p_{l}$, $e^{\Phi(r)}=\left(\frac{r}{r_{0}} \right)^{\beta}$, & L-M & No & No & Yes \\
 & $0< \beta < 2$ & & &  & \\ \hline
$ds^{2}=\left( \frac{\varrho}{\varrho_{0}} \right)^{2\beta}dt^{2} - \frac{d\varrho^{2}}{\left( \frac{\varrho}{\varrho_{0}} \right)^{-\frac{2\left( \beta^{2}-2\beta -1 \right)}{1+\beta}}  -1} -\left( \frac{\beta^{2}-2\beta -2}{\beta^{2}-2\beta -1} -1 \right)\varrho^{2}d\Omega_{ps}^{2}$, & $p_{r}=p_{l}$, $e^{\Phi(r)}=\left(\frac{r}{r_{0}} \right)^{\beta}$, & L-M & No & No & Yes \\
 & $1-\sqrt{2}< \beta < 0$ or & & &   & $2<\beta<1+\sqrt{2}$  \\
 & $2 < \beta < 1+ \sqrt{2}$ & & &  &  \\ \hline
$ds^{2}=\left( \frac{\varrho}{\varrho_{0}} \right)^{2\beta}dt^{2} - \frac{d\varrho^{2}}{1 - \left( \frac{\varrho}{\varrho_{0}} \right)^{-\frac{2\left( \beta^{2}-2\beta -1 \right)}{1+\beta}}} -\left(1- \frac{\beta^{2}-2\beta -2}{\beta^{2}-2\beta -1} \right)\varrho^{2}d\Omega_{ps}^{2}$, & $p_{r}=p_{l}$, $e^{\Phi(r)}=\left(\frac{r}{r_{0}} \right)^{\beta}$, & M-T & No & No & Yes \\
 & $\beta < 1-\sqrt{3}$ or & & &  & $\beta < -1$ or \\
 & $1+\sqrt{3} < \beta$ & & &  & $1+\sqrt{3} < \beta$ \\ \hline
$ds^{2}=\left( \frac{\varrho}{\varrho_{0}} \right)^{2\beta}dt^{2} - \frac{d\varrho^{2}}{1 - \left( \frac{\varrho}{\varrho_{0}} \right)^{-\frac{2\left( \beta^{2}-2\beta -1 \right)}{1+\beta}}} -\left(1- \frac{\beta^{2}-2\beta -2}{\beta^{2}-2\beta -1} \right)\varrho^{2}d\Omega_{ps}^{2}$, & $p_{r}=p_{l}$, $e^{\Phi(r)}=\left(\frac{r}{r_{0}} \right)^{\beta}$, & M-T & No & No & Yes \\
 & $1+\sqrt{2} < \beta < 1+ \sqrt{3}$ or & & &  & $1+\sqrt{2} < \beta < 1+ \sqrt{3}$ \\
 & $1-\sqrt{3} < \beta < 1-\sqrt{2}$ & & &  &  \\ \hline

\end{tabular}

\caption{In this table all discussed by us wormhole solutions are
listed. We indicate to which class a given solution belongs and
which of energy conditions $\rho + p_{i} \geq 0$, $\rho + p_{i} \geq
0$ and $\rho + p_{r} + 2p_{l} \geq 0$ are satisfied or not. We can
see that there are wormholes of L-M and M-T class which satisfy the
energy condition $\rho + p_{r} + 2p_{l} \geq 0$. Notice that for
wormholes with isotropic pressure the condition $\rho + p_{r} +
2p_{l} \geq 0$ becomes $\rho + 3p \geq 0$.}

\label{Tabla1}
\end{table}
\endgroup
\end{center}

\end{widetext}

\subsubsection{Wormholes without angle deficit or excess}
As we have stated above the solution~(\ref{2 metric 1AA}) is a
spacetime carrying a topological defect. We are interested first in
considering pseudo-spherical wormhole geometries without such a
topological defect. Hence we must require for the metric~(\ref{2
metric 1AA}) that
\begin{eqnarray}
1-{\frac {\beta^2- 2\beta-2 }{\beta^2-2\,\beta-1}}=\pm 1.
\end{eqnarray}
Therefore, we have $\beta=1\pm \sqrt{3}$ for the upper sign, and
$\beta=0,2$ for the lower sign.

Let us now consider the values $\beta=1\pm \sqrt{3}$. In this case
the solutions are given by
\begin{eqnarray}\label{metric sin deficit}
ds^2=\left( \frac{r}{r_0}  \right)^{2(1\pm \sqrt{3})}
dt^2-\frac{dr^2}{1-\left( \frac{r_0}{r} \right)^{2/(2\pm
\sqrt{3})}}-
\nonumber \\
r^2 d\Omega_{ps}^{2}, \\
\kappa \rho = \frac{r_0^4(\pm 2 \sqrt{3}-3)}{r^6 \left(\frac{r_0}{r} \right)^{\pm 2 \sqrt{3}}}-\frac{2}{r^2}, \\
\kappa p =  \frac{r_0^4(\mp 2 \sqrt{3}-3)}{r^6 \left(\frac{r_0}{r}
\right)^{\pm 2 \sqrt{3}}}+\frac{2(2 \pm \sqrt{3})}{r^2},
\end{eqnarray}
where we have put, without any loss of generality, $C=1$. Note that
Eq.~(\ref{metric sin deficit}) implies that $g_{rr} \rightarrow 1$
if $r \rightarrow \infty$, and then the M-T constraint 3 is
satisfied. Also these solutions satisfy the flare-out condition
$\frac{b-b^\prime r}{2b^2}>0$, which in this case is clearly
satisfied since
\begin{eqnarray}
\frac{b-b^\prime r}{2b^2}=\frac{\left( \frac{r_0}{r}
\right)^{-\frac{2}{2\pm\sqrt{3}}}}{r(2\pm\sqrt{3})}>0,
\end{eqnarray}
for any $r \geq r_0$.

On the other hand, the case $\beta=0$ was already discussed (see
Eq.~(\ref{WH prho3})). For $\beta=2$ we obtain the solution
\begin{eqnarray}\label{HWH15}
ds^2=\left( \frac{r}{r_0}  \right)^{4} dt^2-\frac{dr^2}{\left(
\frac{r}{r_0} \right)^{2/3}-1}-
r^2 d\Omega_{ps}^{2}, \nonumber \\
\kappa \rho=-\frac{5}{3 r_0^2} \, \left(\frac{r_0}{r} \right)^{4/3},
\,\,\,\,\,\,\,\,\,\,\,\,\,\,\,\,\,\,\,\,\,\,\,\,\,\,\,\,\,\,
\nonumber \\ \kappa p=\frac{5}{r_0^2} \, \left(\frac{r_0}{r}
\right)^{4/3}-\frac{4}{r^2}.
 \,\,\,\,\,\,\,\,\,\,\,\,\,\,\,\,\,\,\,\,\,\,\,\,\,\,\,\,\,\,
 \end{eqnarray}
where we have put $C=-1$ in order to have the right signature in the
metric.

In conclusion these four solutions without a topological defect
belong to two different classes of wormholes: solutions~(\ref{WH
prho3}) and~(\ref{HWH15}) belong to the hyperbolic class of
wormholes discussed by Lobo and Mimoso in Ref.~\cite{Lobo HWH}, and
they clearly do not satisfy the M-T constraint 3, while the
spacetimes~(\ref{metric sin deficit}) belong to the M-T wormhole
class.

\subsubsection{Hyperbolic L-M wormholes with an angle deficit or excess}
We now provide a physical/geometric interpretation of the obtained
pseudo-spherical spacetime~(\ref{2 metric 1AA}), by considering the
presence of a topological defect in the metric~(\ref{2 metric 1AA})
for any $r \geq r_0$. It can be shown that this metric has space
sections with an angle deficit (or excess) in the case of L-M
wormhole class as well as of M-T one.

For having a wormhole configuration we must require $b(r_0)=r_0$.
Then, the metric~(\ref{2 metric 1AA}), energy density and pressure
are provided by the expressions
\begin{widetext}
\begin{eqnarray}\label{metric 15}
ds^2&=&\left(\frac{r}{r_0} \right)^{2 \beta} dt^2-
\frac{dr^2}{\left({\frac {\beta^2- 2\beta-2
}{\beta^2-2\,\beta-1}-1}\right)\left({\left(\frac{r}{r_0}
\right)}^{-{\frac {2(\beta^2-2 \beta-1)}{1+\beta}}}-1\right)}-
r^2 d\Omega_{ps}^{2},  \\
\kappa \rho(r)&=&-
\frac{(2\beta+1)(\beta-3)}{(1+\beta)(\beta^2-2\beta-1) \, r_0^2} \,
\left(\frac{r_0}{r}\right)^{\frac{2\beta(\beta-1)}{1+\beta}}-{\frac
{ \left( \beta-2 \right) \beta}{ \left(
-2\,\beta-1+{\beta}^{2} \right) {r}^{2} }}, \label{pr 15}  \\
\kappa p(r)&=&- \frac{2\beta+1}{(\beta^2-2\beta-1)  r_0^2} \,
\left(\frac{r_0}{r}\right)^{\frac{2\beta(\beta-1)}{1+\beta}}+{\frac
{ \beta^2}{ \left( \beta^2-2\,\beta-1 \right) {r}^{2} }}, \label{pl
15}
\end{eqnarray}
\end{widetext}
respectively. It becomes clear from Eqs.~(\ref{pr 15}) and~(\ref{pl
15}) that the fluid density and the isotropic pressure are not
related by a linear equation of state.

We have such a linear equation of state only for $\beta=0$ and
$\beta=-1/2$. For vanishing $\beta$ we obtain
$\rho/3=-p=-\frac{1}{\kappa r_0^2}$, i.e. $\omega=-1/3$. This
solution is the discussed above zero-tidal-force L-M
wormhole~(\ref{WH prho3}). For $\beta=-1/2$ we obtain for the
metric, energy density and pressure that
\begin{eqnarray}
ds^2&=&\frac{r_0}{r}
dt^2-\frac{dr^2}{4\left(1-\frac{r_0}{r}\right)}-r^2 d
\Omega_{ps}^{2}\\
\kappa \rho&=&-\frac{5}{r^2}, \label{15AAA}\\
\kappa p&=&  \frac{1}{ r^2}. \label{15AAAA}
\end{eqnarray}
As we will discuss below, such a metric represent a M-T hyperbolic
wormhole exhibiting an angle excess.

Let us first show that the metric~(\ref{2 metric 1AA}) includes, as
particular solutions, pseudo-spherical spacetimes of L-M wormhole
class carrying a topological defect. The metric~(\ref{metric 15})
describes L-M pseudo-spherical wormholes if $\frac {(\beta^2-2
\beta-1)}{1+\beta}<0$, i.e. for $1-\sqrt{2}<\beta<1+\sqrt{2}$. Note
that at this range ${\frac {\beta^2- 2\beta-2
}{\beta^2-2\,\beta-1}-1}>0$. By making, for the radial coordinate,
the rescaling
\begin{equation}
r=\pm \sqrt{\frac{\beta^2-2\beta-2}{\beta^2-2\beta-1}-1} \, \varrho,
\end{equation}
the metric~(\ref{metric 15}) takes the form
\begin{eqnarray}\label{metric con deficit angular hyp c=-1}
ds^2=\left(\frac{\varrho}{\varrho_0} \right)^{2 \beta} dt^2-
\frac{d\varrho^2}{ \, {\left(\frac{\varrho}{\varrho_0}
\right)}^{-{\frac { 2(\beta^2-2 \beta-1) }{1+\beta}}}-1}- \nonumber \\
\left({\frac {\beta^2 -2 \beta-2 }{\beta^2-2\,\beta-1}}-1 \right)
\varrho^2 d\Omega_{ps}^{2}.
\end{eqnarray}

The condition $-\frac{1}{{\beta}^{2}-2\,\beta-1 }={\frac {\beta^2 -2
\beta-2 }{\beta^2-2\,\beta-1}}-1>0$ ensures the right signature of
the metric~(\ref{metric con deficit angular hyp c=-1}). For
$1<\frac{\beta^2-2\beta-2}{\beta^2-2\beta-1}<2$, i.e. for $0< \beta
< 2 $, we have a spacetime with an angle deficit, and for
$\frac{\beta^2-2\beta-2}{\beta^2-2\beta-1}>2$, i.e. for
$1-\sqrt{2}<\beta<0$ or $2<\beta < 1+\sqrt{2}$, we have a spacetime
with an angle excess. Notice that the power of $\varrho/\varrho_0$
in the radial metric component of Eq.~(\ref{metric con deficit
angular hyp c=-1}) is positive for $0< \beta < 2 $, so there exist
wormhole of L-M class with an angle deficit. For
$1-\sqrt{2}<\beta<0$ and $2<\beta < 1+\sqrt{2}$ the power $-{\frac {
2(\beta^2-2 \beta-1) }{1+\beta}}$ is also positive, so also there
exist wormholes of the L-M class with an angle excess. In this case
the energy density and pressure are given by
\begin{eqnarray}
\kappa \rho(\varrho)&=& \frac{(2\beta+1)(\beta-3)}{(1+\beta)
\varrho_0^2} \,
\left(\frac{\varrho}{\varrho_0}\right)^{-\frac{2\beta(\beta-1)}{1+\beta}}+{\frac
{ \left( \beta-2 \right) \beta}{ {\varrho}^{2} }}, \nonumber \\
\kappa p(\varrho)&=& \frac{(2\beta+1)}{\varrho_0^2} \,
\left(\frac{\varrho}{\varrho_0}\right)^{-\frac{2\beta(\beta-1)}{1+\beta}}-{\frac
{ \beta^2}{ {\varrho}^{2} }}. \label{915}
\end{eqnarray}
It is interesting to note, that for ${\frac {\beta^2-2
\beta-1}{1+\beta}}<0$, the wormhole condition $b(r_0)=r_0$ has
induced for the metric~(\ref{metric 15}) the behavior $g_{rr}^{-1}
\rightarrow \frac{1}{1+2\beta-\beta^2} \, (r/r_0)^{\left|2{\frac
{(\beta^2-2 \beta-1)}{1+\beta}}\right|}$ for $r \rightarrow \infty$,
i.e. the asymptotic spacetime exhibits an angle deficit (or excess).

\begin{figure}
\includegraphics[scale=0.3]{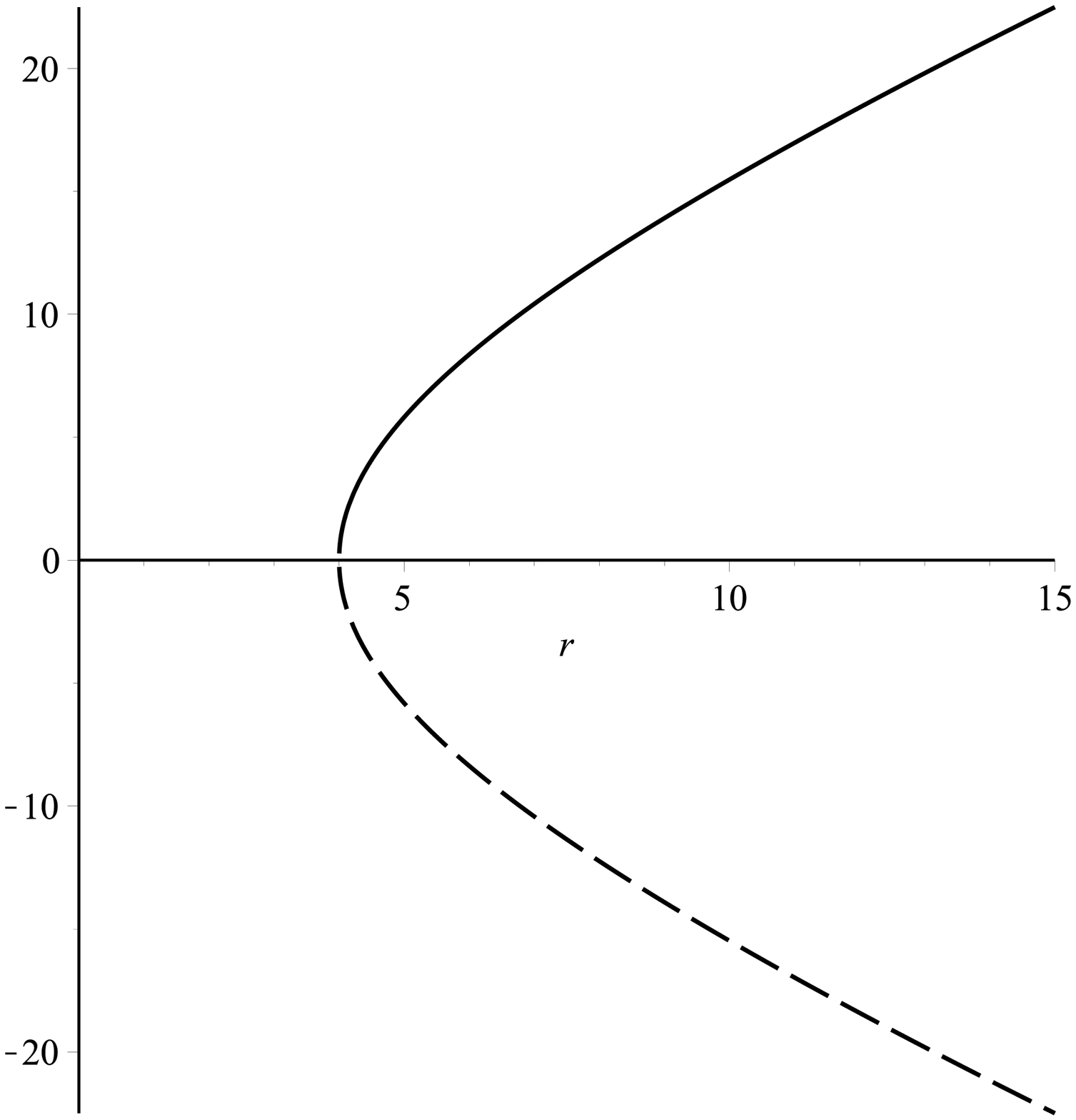}
\caption{The figure shows the embedding diagram of the
wormhole~(\ref{caso particular hwh}) obtained from Eq.~(\ref{ED15})
for the parameter values $\beta=3, r_0=4$, where the throat is
located. The solid line correspond to $+$ sign and the dashed line
to $-$ sign in Eq.~(\ref{ED15}). The embedding shows that the
wormhole extends from the throat at $r=4$ to infinity. For a full
visualization of the surface the diagram must be rotated about the
vertical z axis (see Fig.~(\ref{Fig15BB15})). } \label{Fig15AA}
\end{figure}

\begin{figure}
\includegraphics[scale=0.3]{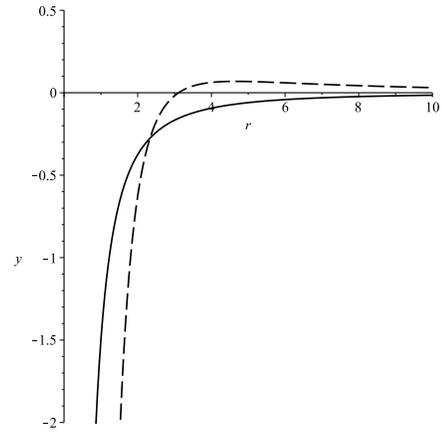}
\caption{The figure shows the behavior of the energy density (solid
line) and the pressure (dashed line). The energy density is always
negative, while the pressure is always positive for $r\geq 4$ (note
that p=0 at $r=28/9<4$, where the throat is  located).}
\label{Fig15BB}
\end{figure}

\begin{figure}
\includegraphics[scale=0.5]{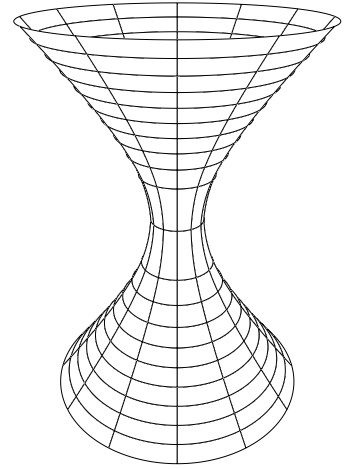}
\caption{The figure shows the three-dimensional embedding diagram
for the wormhole of Fig.~(\ref{Fig15AA}).} \label{Fig15EAA}
\label{Fig15BB15}
\end{figure}

\subsubsection{Hyperbolic M-T wormholes with an angle deficit or excess}
In order to metric~(\ref{metric 15}) describes solutions of the M-T
wormhole class the M-T Constraint 3  must be satisfied. Hence we
must require that ${\frac {\beta^2-2 \beta-1}{1+\beta}}>0$, implying
that the parameter $\beta$ varies in the ranges $-1<\beta<
1-\sqrt{2}$ or $1+\sqrt{2}<\beta$. In this case the
metric~(\ref{metric 15}) may be rewritten in the equivalent form
\begin{eqnarray}\label{metric 915}
ds^2=\left(\frac{r}{r_0} \right)^{2 \beta} dt^2- \nonumber \\
\frac{dr^2}{\left({1-\frac {\beta^2- 2\beta-2
}{\beta^2-2\,\beta-1}}\right)\left(1-{\left(\frac{r}{r_0}
\right)}^{-{\frac {2(\beta^2-2 \beta-1)}{1+\beta}}}\right)}- r^2
d\Omega_{ps}^{2}.
\end{eqnarray}
If $r \rightarrow \infty$ we obtain $g_{rr}^{-1} \rightarrow
1-{\frac {\beta^2- 2\beta-2 }{\beta^2-2\,\beta-1}} $, implying that
the asymptotic spacetime is given by the metric~(\ref{metric 15 DA
15}), which describes a space with an angle deficit (or excess). By
making, for the radial coordinate, the rescaling
\begin{equation}
r=\pm \sqrt{1-\frac{\beta^2-2\beta-2}{\beta^2-2\beta-1}} \, \varrho,
\end{equation}
the metric~(\ref{metric 15}) takes the form
\begin{eqnarray}\label{metric con deficit angular hyp}
ds^2=\left(\frac{\varrho}{\varrho_0} \right)^{2 \beta} dt^2-
\frac{d\varrho^2}{1- {\left(\frac{\varrho}{\varrho_0}
\right)}^{-{\frac { 2(\beta^2-2 \beta-1) }{1+\beta}}}}- \nonumber \\
\left(1-{\frac {\beta^2 -2 \beta-2 }{\beta^2-2\,\beta-1}} \right)
\varrho^2 d\Omega_{ps}^{2}.
\end{eqnarray}

The condition $\frac{1}{{\beta}^{2}-2\,\beta-1 }=1-{\frac {\beta^2
-2 \beta-2 }{\beta^2-2\,\beta-1}}>0$ must be required in order to
ensure the right signature of the metric~(\ref{metric con deficit
angular hyp}). For $0<\frac{\beta^2-2\beta-2}{\beta^2-2\beta-1}<1$,
i.e. for $\beta
> 1+\sqrt{3}$ or $\beta < 1-\sqrt{3}$, we have a spacetime with an
angle deficit, and for
$\frac{\beta^2-2\beta-2}{\beta^2-2\beta-1}<0$, i.e. for
$1+\sqrt{2}<\beta<1+\sqrt{3}$ or $1-\sqrt{3}<\beta < 1-\sqrt{2}$, we
have a spacetime with an angle excess. Notice that the power of
$\varrho/\varrho_0$ in the radial metric component of
Eq.~(\ref{metric con deficit angular hyp}) is negative for $\beta
> 1+\sqrt{3}$ and $-1<\beta < 1-\sqrt{3}$,
so there exist wormhole of M-T class with an angle deficit. For
$1+\sqrt{2}<\beta<1+\sqrt{3}$ and $1-\sqrt{3}<\beta < 1-\sqrt{2}$
the power $-{\frac { 2(\beta^2-2 \beta-1) }{1+\beta}}$ is also
negative, so also there exist wormholes of the M-T class with an
angle excess. In this case the energy density and pressure are given
by
\begin{eqnarray}
\kappa \rho(\varrho)&=&- \frac{(2\beta+1)(\beta-3)}{(1+\beta)
\varrho_0^2} \,
\left(\frac{\varrho}{\varrho_0}\right)^{-\frac{2\beta(\beta-1)}{1+\beta}}-{\frac
{ \left( \beta-2 \right) \beta}{ {\varrho}^{2} }}, \nonumber \\
\kappa p(\varrho)&=&- \frac{(2\beta+1)}{\varrho_0^2} \,
\left(\frac{\varrho}{\varrho_0}\right)^{-\frac{2\beta(\beta-1)}{1+\beta}}+{\frac
{ \beta^2}{ {\varrho}^{2} }}. \label{915A}
\end{eqnarray}
We note that expressions~(\ref{915}) and~(\ref{915A}) for the energy
density and isotropic pressures are identical up to the sign.

Now, in order for the spacetime~(\ref{metric 15}) to see the shape
of a wormhole of the hyperbolic M-T class, we will embed space
slices of the metric~(\ref{metric 15}), or equivalently of the
metric~(\ref{metric 915}), satisfying the conditions ${\frac
{\beta^2-2 \beta-1}{1+\beta}}>0$ and $1-{\frac {\beta^2- 2\beta-2
}{\beta^2-2\,\beta-1}}>0$, as surfaces of revolution in an Euclidean
3-dimensional space. For producing embeddings we shall consider two
dimensional space slices of the metric~(\ref{metric 15}) by making
the restriction $t=t_0=const$ and $\sinh(\theta_0)=1$ (which implies
that $\theta_0=\ln(1\pm \sqrt{2})$). With these constraints the
respective two-dimensional line element may be written as
\begin{eqnarray}\label{2 metric 15}
ds^2=\frac{dr^2}{\left({1-\frac {\beta^2- 2\beta-2
}{\beta^2-2\,\beta-1}}\right)\left(1-{\left(\frac{r}{r_0}
\right)}^{-{\frac {2(\beta^2-2 \beta-1)}{1+\beta}}}\right)}+ r^2
d\phi^2, \nonumber \\
\end{eqnarray}
To visualize this space slice we identify Eq.~(\ref{2 metric 15}) as
the metric of a surface of revolution in $R^3$, embedded into
Euclidean metric, written in cylindrical coordinates as
\begin{eqnarray}
ds^2=dz^2+dr^2+r^2 d\phi^2.
\end{eqnarray}
Comparing both metrics we conclude that
\begin{eqnarray}\label{ED15}
\frac{dz}{dr}=\pm
\frac{1}{\sqrt{\frac{\beta^2-2\beta-1}{\beta^2-2\beta-2+\left(\frac{r_0}{r}
 \right)^{\frac{2(\beta^2-2\beta-1)}{1+\beta}}}-1}}.
\end{eqnarray}
Let us briefly discuss particular solutions defined by the parameter
values $\beta=3, r_0=4$ and $\beta=-1/2, r_0=~4$. For $\beta=3$ and
$r_0=4$ the respective metric, energy density and pressure are given
by
\begin{eqnarray}\label{caso particular hwh}
ds^2&=&\frac{r^2}{64} \,
dt^2-\frac{dr^2}{\frac{1}{2}-\frac{2}{r}}-r^2 d\Omega_{ps}^{2}, \\
\kappa \rho&=&-\frac{3}{2r^2}, \\
\kappa p&=& -\frac{14}{r^3}+\frac{9}{2 r^2}.
\end{eqnarray}
This spacetime describes an hyperbolic wormhole with a throat
located at $r=4$, and in the asymptotic limit we obtain a spacetime
with an angle deficit. In this case, the coordinates $\theta$ and
$\phi$ are redefined as follows: \\
$-\infty<\tilde{\theta}=\theta/\sqrt{2}<\infty$ and $0 \leq
\tilde{\phi}=\phi/\sqrt{2} \leq \sqrt{2} \pi$.

In Fig.~(\ref{Fig15AA}) we show the embedding of this hyperbolic
wormhole, while in Fig.~(\ref{Fig15BB}) the behavior of the energy
density and the isotropic pressure are shown.

\begin{figure}
\includegraphics[scale=0.3]{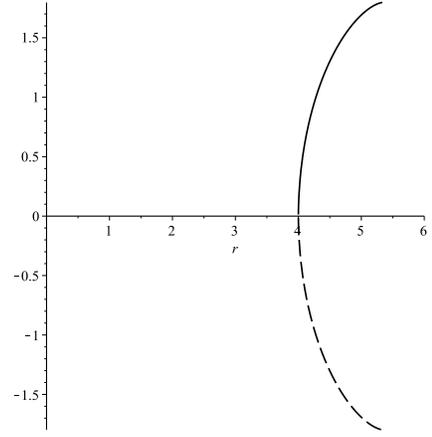}
\caption{The figure shows the embedding diagram of the
wormhole~(\ref{caso particular hwh154}) obtained from
Eq.~(\ref{ED15}) for the parameter values $\beta=-1/2, r_0=4$, where
the throat is located. The solid line correspond to $+$ sign and the
dashed line to $-$ sign in Eq.~(\ref{ED15}). In this case the
wormhole extends from the throat, at $r=4$, to $r=16/3$. For a full
visualization of the surface the diagram must be rotated about the
vertical z axis (see Fig.~(\ref{Fig15EAB})).} \label{Fig15EA}
\end{figure}

Now, for $\beta=-1/2$ and $r_0=4$ the metric is provided by
\begin{eqnarray}\label{caso particular hwh154}
ds^2&=&\frac{1}{r} \, dt^2-\frac{dr^2}{4-\frac{16}{r}}-r^2
d\Omega_{ps}^{2},
\end{eqnarray}
while the energy density and pressure are given by
Eqs.~(\ref{15AAA}) and ~(\ref{15AAAA}) respectively. This spacetime
describes an hyperbolic wormhole with a throat located at $r=4$, and
in the asymptotic limit we obtain a spacetime with an angle excess.
The coordinates $\theta$ and $\phi$ are redefined as follows:
$-\infty<\tilde{\theta}=2\theta<\infty$ and $0 \leq
\tilde{\phi}=2\phi  \leq 4 \pi$. The embedding shows (see
Fig.~(\ref{Fig15EA})) that the wormhole extends from the throat, at
$r=4$, to $r=16/3$.

\begin{figure}
\includegraphics[scale=1.5]{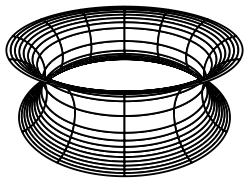}
\caption{The figure shows the three-dimensional embedding diagram
for the wormhole of Fig.~(\ref{Fig15EA}).} \label{Fig15EAB}
\end{figure}

\section{Conclusions}
We have shown that in pseudo-spherically symmetric spacetimes there
are two classes of static wormhole solutions: one of them discussed
by Lobo and Mimoso in Ref.~\cite{Lobo HWH}, and another one
discussed by us in this paper. The L-M class may be written in the
form~(\ref{LM BH metric}) and is characterized by the
condition~(\ref{condition lm}), while the second class of hyperbolic
wormholes may be written in the form~(\ref{general pseudospherically
BH metric}) and is characterized by the M-T condition~(\ref{br1}).
The use of Eqs.~(\ref{br1}), (\ref{LM BH metric}), (\ref{condition
lm}) and~(\ref{general pseudospherically BH metric}) is the main
criterion to determine to which class of wormholes a given
hyperbolic solution belongs.

We obtain exact solutions belonging to both classes of static and
pseudo-spherically symmetric spacetime wormholes. The specific
studied solutions are obtained by considering several equations of
state or by imposing restricted choices for the redshift function
and/or the shape function.

It is interesting to remark that for all obtained pseudo-spherical
wormhole solutions we have that at the throat the energy density is
negative, while the radial pressure positive. In general, it can be
shown that this is true for any hyperbolic wormhole, including L-M
wormholes, as well as M-T ones. Effectively, from the radial metric
component in metric~(\ref{general pseudospherically BH metric}) we
may write for the proper radial distance that
\begin{eqnarray}
l(r)=\pm \int_{r_0}^{r} \frac{d
r^\prime}{\sqrt{1-\frac{b(r^\prime)}{r^\prime}}}.
\end{eqnarray}
From this expression it can be shown that at the throat always the
relation
\begin{eqnarray}\label{ct15}
b^\prime(r_0) \leq 1
\end{eqnarray}
is satisfied~\cite{Morris:1988tu}. Thus, from Eq.~(\ref{p rho}) we
obtain for the energy density $\kappa \rho(r_0) \leq -1/r_0^2$,
implying that at the throat the energy density is always negative.

On  the other hand, if the redshift function is finite for $r \geq
r_0$, Eq.~(\ref{p pr}) implies that $\kappa p_r(r_0)=1/r_0^2$, so
the radial pressure is always positive at the throat. Notice that
$\kappa \rho+\kappa p_r|_{r_0}= (b^\prime(r_0)-1)/r_0^2$. Thus, by
taking into account Eq.~(\ref{ct15}) we conclude that at the throat
$\rho+p_r|_{r_0}\leq0$.

From Eq.~(\ref{p pl}) we obtain that at the throat the relation
$\kappa p_l(r_0)=(1-b^\prime)(1+\phi^\prime(r_0) r_0+1)/(2r_0^2)$ is
valid for the lateral pressure. Therefore, Eq.~(\ref{ct15}) implies
that, at the throat, for $\phi^\prime(r_0)> -1/r_0$ the lateral
pressure is positive, while for $\phi^\prime(r_0)< -1/r_0$, the
lateral pressure is negative. This is why for the zero-tidal-force
hyperbolic wormholes~(\ref{metricalobo}) and~(\ref{ztf hwh}) the
lateral pressures~(\ref{presionlobo}) and~(\ref{plhwh}) are
positive. In conclusion, it follows from these results that an
hyperbolic wormhole can not be sustained at the throat by phantom
energy, since always the energy density is negative, and at least
the radial pressure is always positive.

Notice that in opinion of the authors of~\cite{Lobo HWH} the static
hyperbolic tunnels are constructed by adding exotic matter to the
vacuum solution~(\ref{VSS}), which is the pseudo-spherical vacuum
counterpart of the Schwarzschild solution.

On the light of our results, this statement is not true, since,
strictly speaking, the metric~(\ref{VSS}) should be obtained from
any hyperbolic wormhole when the exotic matter vanishes (we should
obtain a background solution with $g_{tt}=2 \mu/r-1$, which vanishes
at $r=2 \mu$). This clearly does not happen, even in the case of
hyperbolic L-M wormholes discussed in Ref.~\cite{Lobo HWH}.

In our case, most of static hyperbolic M-T wormholes are constructed
by adding exotic matter to the spacetime~(\ref{hip}). This spacetime
is the hyperbolic counterpart of the Minkowski background (and not
of the vacuum Schwarzschild solution), on which most of spherically
symmetric wormholes are constructed by adding exotic matter to it.

For example, for the zero-tidal-force hyperbolic M-T
wormhole~(\ref{ztf hwh}) we have that if $r_0=0$ the background
solution~(\ref{hip}) is obtained. While, for the zero-tidal-force
hyperbolic L-M wormhole~(\ref{metricalobo}), neither the
metric~(\ref{hip}) nor~(\ref{VSS}) are the background spacetimes on
which the solution~(\ref{metricalobo}) is constructed by adding
exotic matter to one of them.

In the table~\ref{Tabla1} all discussed by us wormhole solutions are
listed. We indicate to which class a given solution belongs and
which energy conditions are satisfied.

Lastly, let us remark that the counterpart of the static and
spherical self-dual Lorentzian wormhole discussed in
Ref.~\cite{Dadhich} is not a wormhole in pseudo-spherically
symmetric spacetimes. Effectively, this two-parameter family of
spherically symmetric, static Lorentzian wormholes is obtained as
the general solution of the equations $\rho(r)=0$ and
$T_{ij}-\frac{1}{2} T g_{ij} u^i u^j=0$, where $u^i u_j=1$. All
these solutions have a vanishing scalar curvature $R = 0$. The
conditions required for these self-dual Lorentzian wormholes imply
that in pseudo-spherical spacetimes the metric is given by
\begin{eqnarray*}
ds^2=\left( k+\lambda \sqrt{\frac{2m}{r}-1} \right)^2
dt^2-\frac{dr^2}{\frac{2m}{r}-1}-   r^2 d\Omega_{ps}^{2},
\end{eqnarray*}
where $k$, $\lambda$ and $m$ are constants of integration. This
class of solutions, as well as the spherical symmetric one, includes
the vacuum hyperbolic version of the spherically symmetric
Schwarzschild solution, which is obtained by requiring $k=0$. This
solution is obtained by adding a matter source to the vacuum
metric~(\ref{VSS}), however, in order to have $g_{rr}
>0$ we must require that $0<r<2m$, and then this spacetime does not
describe a wormhole geometry.

{\bf Acknowledgements:} This work was supported by Direcci\'on de
Investigaci\'on de la Universidad del B\'\i o-B\'\i o through grants
N$^0$ DIUBB 140708 4/R and N$^0$ GI 150407/VC.

\end{document}